\documentstyle[12pt]{article}
\normalsize

\def\bar#1{\overline{#1}}

\def\Hat#1{\rlap{\kern.10em$\widehat{\phantom G}$}#1}
\def\HAt#1{\rlap{\kern.05em$\widehat{\phantom G}$}#1}

\def\cap#1{\rlap{\kern.1em$\widehat{\phantom{G\vrule height.8em}}$}#1{}}
\def\Cap#1{\rlap{\kern.05em$\widehat{\phantom{G\vrule height.8em}}$}#1{}}

\newcounter{sxn}

\newcounter{axn}

\def\br{\begin{thebibliography}{abc}}
\def\er{\end{thebibliography}}

\date{}

\begin{document}
\bibliographystyle{unsrt}
\footskip 1.0cm
\thispagestyle{empty}
\setcounter{page}{0}
\begin{flushright}
SU-4240-537\\
September 1993\\
\end{flushright}
\vspace{10mm}

\centerline {\LARGE INEQUIVALENT QUANTIZATIONS OF YANG-MILLS}
\vspace{5mm}
\centerline {\LARGE THEORY ON A CYLINDER}
\vspace*{15mm}
\centerline {\large L. Chandar and E. Ercolessi}

\vspace*{5mm}
\centerline {\it Department of Physics, Syracuse University,}
\centerline {\it Syracuse, NY 13244-1130, U.S.A.}
\vspace*{25mm}
\normalsize
\centerline {\bf Abstract}
\vspace*{5mm}

Yang-Mills theories on a 1+1 dimensional cylinder are considered. It is shown
that canonical quantization can proceed following different routes, leading to
inequivalent quantizations.

The problem of the non-free action of the gauge group on the configuration
space is also discussed. In particular we re-examine the relationship between
``$\theta$-states" and the fundamental group of the configuration space. It is
shown that this relationship does or does not hold depending on whether or not
the gauge transformations not connected to the identity act freely on the
space of connections modulo connected gauge transformations.

\newpage

\baselineskip=24pt
\setcounter{page}{1}
\newcommand{\be}{\begin{equation}}
\newcommand{\ee}{\end{equation}}

To the present day, the understanding of the canonical quantization of
non-abelian gauge theories in 3+1 dimensions and the knowledge of their
physical
degrees of freedom are lacking the necessary rigor.  Recent papers \cite{1,2}
have dealt with the somewhat simpler task of quantizing $SU(N)$ or $U(N)$
Yang-Mills theory on a 1+1 dimensional cylinder.

These cases are interesting not only because they may cast some light on the
features of the 3+1 dimensional case, but also because they deal with the
quantization of a gauge theory on a compact spatial manifold.  Indeed, it is
known \cite{3} that on compact spaces, the action of the gauge group on the
space of connections may not be free, causing the phase space to be no longer a
manifold but only an orbifold.  In relation to both Yang-Mills theories and
gravity, some simplified finite-dimensional examples have been worked out in
\cite{5} and \cite{6}.

In this paper, we describe some novel features that arise in the quantization
of pure Yang-Mills theory on a cylinder.
It will be shown that this model admits inequivalent quantizations, the
ambiguities arising because of two different reasons:
\begin{enumerate}
   \item Let ${\cal G}$ denote the gauge group, i.e. the group of all gauge
transformations that act on the phase space, and ${\cal G}_{0}$ the subgroup of
${\cal G}$ connected to the identity, which is generated by Gauss law.  As in
any
gauge theory, ${\cal G}_{0}$ should leave a physical state invariant.
Towards this end, one will have to go to a reduced phase space and/or impose
Gauss law on the physical states.  In the problem at hand it will be shown that
there are inequivalent ways of carrying out this procedure.
    \item In our problem, the action of the gauge group is not free and the
reduced phase space is an orbifold.  After suitably treating the singular
points \cite{5}, it will be shown that there are different self-adjoint
extensions of the relevant operators, and hence different quantum theories.
\end{enumerate}

The fact that the action of the gauge group is not free has also another
consequence. In non-abelian gauge theories in 3+1 dimensions, the possibility
of inequivalent quantizations and $\theta$-states arises because of the
existence of gauge transformations that are not connected to the identity (i.e.
elements of ${\cal G}^{\infty}$ which are not elements of its identitity
component ${\cal G}^{\infty}_{0}$, the superscript $\infty$ indicating that the
gauge transformations have to vanish at spatial infinity). It is known \cite{4}
that this is closely related to the multiple connectivity of the reduced
configuration space. But the usual arguments relating them require a
\underline{free} action of the group ${\cal G}^{\infty}$ on the space ${\cal
A}$ of connections. Indeed, if this action is free one can show that the
fundamental group $\Pi_{1}({\cal Q})$ of the reduced configuration space
${\cal Q}={\cal A}/{\cal G}^{\infty}$ equals the group of disconnected
components of ${\cal G}^{\infty}$, given by ${\cal G}^{\infty}/{\cal
G}^{\infty}_{0}$. That this action is free can be proven whenever the gauge
transformations in question are \underline{pointed} gauge tranformations, i.e.
they reduce to identity at a point of the base space (as it happens at the
point at $\infty$ in the case of ${\bf R}^{4}$).
Now in the context of gravity, \cite{6} gives an example in which this
relationship
does not hold because the action of the group of gauge transformations
(diffeomorphisms in this case) is not free. In particular it is proven that the
configuration space is simply connected despite the existence of gauge
transformations not connected to the identity.

Let us make a few initial remarks about this problem in our context.

We will consider here pure YM theories when the group $G$ is $U(N)$ or $SU(N)$
and the spatial manifold is $S^{1}$. In all these cases the action of the gauge
group ${\cal G}$ on the space of connections ${\cal A}$ is \underline{never
free}.

For $G=SU(N)$, ${\cal G}/{\cal G}_{0}$ is trivial and therefore there exist no
$\theta$-states. Neither is the reduced configuration space multiply
connected, so that there is nothing much to say about the relation between
$\theta$-states and the fundamental group of the reduced configuration space in
this case.

For $G=U(N)$, we do have gauge transformations not connected to the identity
and
${\cal G}/{\cal G}_{0}={\bf Z}$. As already mentioned above, the action of the
gauge group ${\cal G}$ on the space ${\cal A}$ of connections is not free.
However the group ${\cal G}/{\cal G}_{0}$ of transformations not
connected to the identity still acts freely on the space ${\cal A}/{\cal
G}_{0}$
of connections modulo connected gauge transformations. This is because of the
following reason. [Below we denote the Lie algebra of $G$ by $\underline{G}$]:

If $g(x)$ is an element of ${\cal G}$ which is not connected
to the identity, then its action on a gauge potential $A(x)$ is
\be
A(x) \longrightarrow g(x) \cdot A(x) := g(x) A(x) g(x)^{-1} + g(x) dg(x)^{-1}
\;
{}.
\ee
In particular $A_{U(1)}$, the component of $A$ along the \underline{$U(1)$}
central subalgebra of \underline{$U(N)$}, transforms according to
\be
A_{U(1)} \longrightarrow (g \cdot A)_{U(1)} = A_{U(1)} + (gdg^{-1})_{U(1)} \; ,
\ee
where $(gdg)^{-1})_{U(1)}$ is a closed, non-exact form because the non-trivial
map $g(x)$ is homotopic to a map which necessarily winds around the $U(1)$
center of $U(N)$. This shows that ${\cal G}/{\cal G}_{0}$ acts freely on ${\cal
A}$. That its action is free also on ${\cal A}/{\cal G}_{0}$ follows because
otherwise we would have
\be
g \cdot A = g_{0} \cdot A \Leftrightarrow (g_{0}^{-1}g) \cdot A = A \;\;\;
\mbox{for } g \in {\cal G}/{\cal G}_{0} \;\;\; \mbox{and } g_{0} \in
{\cal G}_{0} \; ,
\ee
which cannot be since $g_{0}^{-1}g$ is also an element in ${\cal G}/{\cal
G}_{0}$.

 Thus the fundamental group $\Pi_{1}({\cal Q})$ of the reduced
configuration space ${\cal Q}={\cal A}/{\cal G}=({\cal A}/{\cal G}_{0})/({\cal
G}/{\cal G}_{0})$ is once again given by ${\cal G}/{\cal G}_{0}={\bf Z}$. For
such a theory, then, the usual relation between $\theta$-states and the
multiple connectivity of the configuration space continues to hold.

Since in this paper we are looking at pure YM theories (without fermions), when
$G=SU(N)$, the relevant group can be very well said to be not SU(N) but
$SU(N)/{\cal C}(SU(N))$ where ${\cal C}(SU(N))$ is the center of $SU(N)$. This
is because the center of $SU(N)$ is a discrete subgroup and therefore the phase
space variables which are Lie algebra valued, as well as the gauge
transformations of such variables, live only on the quotient of $SU(N)$ by its
discrete center. [This would not have been the case if $G$ were $U(N)$ because
the center of the latter is $U(1)$ and hence not discrete.]

Thus, for example, if $G=SU(2)$, we can alternatively think of a theory with
$G=SU(2)/{\bf Z}_{2}=SO(3)$. In this case it turns out that ${\cal G}/ {\cal
G}_{0}={\bf Z}_{2}$ and $\theta$-states associated to gauge transformations
not connected to the identity \underline{do exist} in this theory. [By
``$\theta$-states" we mean those associated with irreducible representations of
${\cal G}/{\cal G}_{0}$, even though, as here, they are not labelled by an
angle
$\theta$.] However, unlike in the $U(N)$ case, here the reduced configuration
space ${\cal Q}={\cal A}/{\cal G}$ is simply connected: since ${\bf Z}_{2}$
does not act freely on the space of connections modulo connected gauge
transformations, the relationship between $\theta$-states and non-simple
connectivity of the configuration space is spoiled .

We now turn to the demonstration of the results 1 and 2.

 Consider pure YM theory for a semisimple compact group $G$ on a
cylinder.  Let us also assume that space is a circle running from $x=0$ to
$x=2\pi$.

The action is
\begin{equation}
S=-\frac{1}{4}\int Tr(F_{\mu\nu}F^{\mu\nu}) \; ,  \label{1}
\end{equation}
where the curvature tensor F is given by
\be
F_{\mu\nu}=\partial _{\mu}A_{\nu}-\partial _{\nu}A_{\mu} + [A_{\mu},A_{\nu}] \;
. \label{2}
\ee

If $T^{a}$ are the generators of the Lie algebra $\underline{G}$ of $G$, then
a Lie algebra valued field $X$ is defined in terms of its components $X^{a}$ by
\be
X = X^{a} T^{a} \; , \label{3}
\ee
where summation over $a$ is assumed.

The Hamiltonian and Poisson Brackets (PB) obtained from (\ref{1}) are
\be
H=\frac{1}{2}\int Tr(E^{2})\, dx \; , \label{4}
\ee
\begin{eqnarray}
\{ A_{1}^{a}(x),A_{1}^{b}(y) \} = & 0 & =\{ E^{a}(x),E^{b}(y) \} \nonumber \\
\{ A_{1}^{a}(x),E_{1}^{b}(y) \} & = & \delta ^{ab}\delta (x-y) \; . \label{5}
\end{eqnarray}

The Hamiltonian (\ref{4}) is to be complemented by the Gauss law
\be
\frac{\partial E}{\partial x}+[A_{1},E] \approx 0 \; , \label{6}
\ee
which is the generator of gauge transformations
\be
(A_{1}\; ,\; E)\longrightarrow (gA_{1}g^{-1} + g\frac{\partial g^{-1}}{\partial
x}\; ,\; gEg^{-1}) \label{7}
\ee
where $g(x)$ is valued in $G$.

Following Rajeev \cite{2}, in order to isolate the gauge invariant degrees of
freedom, we define variables $S(x)$ and $\tilde{E} (x)$:
\begin{eqnarray}
S(x) & = & {\cal P}exp[-\int _{0}^{x} A_{1}(y)\; dy] \; ,\nonumber \\
\tilde{E} (x) & = & S^{-1}(x)E(x)S(x) \; .            \label{8}
\end{eqnarray}
${\cal P}$ denotes path ordering.

Notice that since $S(2\pi )$ need not to be equal to $S(0)={\bf 1}$, this
redefinition imposes a twisted boundary condition on $\tilde{E} (x)$:
\be
S(2\pi ) \tilde{E} (2\pi )S^{-1}(2\pi ) = \tilde{E} (0) \;. \label{9}
\ee

The Hamiltonian and PB in these variables are
\begin{eqnarray}
H & = & \frac{1}{2}\int Tr(\tilde{E}^{2})\, dx  \; ,\label{10}\\
\{ S(x), S(y)\} & = & 0 \; , \nonumber \\
\{ S(x), \tilde{E}^{a}(y)\} & = & -\theta (x-y)S(x)T^{a} \; , \nonumber \\
\{ \tilde{E}^{a}(x), \tilde{E}^{b}(y)\} & = &
f^{abc}\tilde{E}^{c}(y) \; , \;\;\;\;y < x \; . \label{11}
\end{eqnarray}
where $f^{abc}$ refer to the structure constants of the Lie Algebra
$\underline{G}$.

The gauge transformations now assume the simpler form
\be
(S(x)\; ,\; \tilde{E} (x))\longrightarrow (g(x)S(x)g^{-1}(0)\; ,\;
g(0)\tilde{E} (x)g^{-1}(0))  \; .\label{12}
\ee
Furthermore, using
\be
\frac{dS}{dx} = -A_{1}S \; , \label{13}
\ee
Gauss law becomes
\be
\frac{\partial \tilde{E}}{\partial x} \approx 0  \label{14}
\ee
and can be easily solved:
\be
\tilde{E} (x) = \tilde{E} (0) \; . \label{15}
\ee
Using the equivalence under gauge transformations (\ref{12}) and Gauss law
(\ref{15}), it is easy to see \cite{2} that the reduced phase space
consists of
\begin{eqnarray}
q & = & S(2\pi ) \; \in G , \nonumber \\
p & = & \tilde{E} \; \in \underline{G} \; , \label{16}
\end{eqnarray}
along with the constraint arising from (\ref{9}) and the gauge equivalence
(\ref{12}):
\be
qpq^{-1} = p \; , \label{17}
\ee
\be
(q\; ,\; p)\;\; \sim \;\; (gqg^{-1}\; ,\; gpg^{-1}) \; . \label{18}
\ee
Here $g=g(0)$, $g(x)$ being the gauge transformation.

The Hamiltonian (\ref{10}) and PB's (\ref{11}) can be restricted to the
$(q\; ,\; p)$ space:
\begin{eqnarray}
H & = & \pi Tr(p^{2}) \; , \label{19} \\
\{ q, q\} & = & 0 \; , \nonumber \\
\{ q, p^{a}\} & = & -qT^{a} \; , \nonumber \\
\{ p^{a}, p^{b}\} & = & f^{abc} p^{c} \; , \label{20}
\end{eqnarray}
$p^{a}$ is defined here using rule (\ref{3}).

In addition, it can be checked, using (\ref{20}), that the function $qpq^{-1} -
p$ involved in the constraint (\ref{17}) generates exactly the gauge
transformations (\ref{18}).

The following point about (\ref{17}) is to be
noted. In quantum theory, $qpq^{-1}$ is not well defined unless an ordering of
the operators is also defined. However, using the PB's (\ref{20}) between $q$
and $p^{a}$ and the fact that classical Poisson Brackets are replaced by Lie
Brackets in quantum theory, it can be checked that the difference between the
quantum operators $qp^{a}T^{a}q^{-1}$ and (say) $p^{a}qT^{a}q^{-1}$, both of
which correspond classically to $qpq^{-1}$, is only a constant. Due to this
lucky fact, $qpq^{-1} - p$ is well defined (without having to define an
ordering) even in quantum theory up to a constant. Thus the gauge
transformations generated by this function are independent of the ordering.

Having got the canonical structure of the phase space, we can proceed to
quantize the system. We will argue that one can follow three different routes.

Following Rajeev \cite{2}, the first approach to quantization can be to declare
wave functions as complex functions of $q$ which are
invariant under $g\longrightarrow gqg^{-1}$.
Then the Hamiltonian, being just the quadratic Casimir, has as eigenstates the
characters of the irreducible representations. We have nothing new to add to
this approach.

The remaining two alternatives, developed for the first time in this paper,
insist on carrying  through the reduction of the phase space in stages.
In these approaches, we show how quantization ambiguities arise because of
the two reasons mentioned at the beginning.

The first step in this reduction is the observation that the constraint
(\ref{17})
allows us to choose a
representative $(\tilde{q}\; ,\; \tilde{p})$ for each gauge equivalence class
of the following type \cite{1}:
\begin{eqnarray}
\tilde{q} & \in & \mbox{ Cartan subgroup of $G$, denoted $C_{G}$},\nonumber \\
\tilde{p} & \in & \mbox{ Cartan subalgebra of $\underline{G}$, denoted
$\underline{C_{G}}$.} \label{21}
\end{eqnarray}

The consequence is that (\ref{19}) and (\ref{20}) can now be restricted to
the (partially) reduced phase space $C_{G} \times \underline{C_{G}}$, this
restriction making sense because of the gauge invariance of the Hamiltonian and
the gauge covariance of the PB relations. [The meaning of the latter s that
(\ref{20}) is invariant in form under a gauge transformation $q \rightarrow
gqg^{-1}$, $p \rightarrow gpg^{-1}$.] We thus have
\begin{eqnarray}
H & = & \pi
Tr(\tilde{p}^{2}) \; , \label{22} \\
\{ \tilde{q}, \tilde{q}\} & = & 0 \; , \nonumber \\
\{ \tilde{p}^{i}, \tilde{p}^{j}\} & = & 0 \; , \nonumber \\
\{ \tilde{q},\tilde{p}^{i}\} & = & -\tilde{q}C^{i} \; , \label{23}
\end{eqnarray}
(where $i$ runs over the Cartan Subalgebra generators and $C^{i}$ are the
corresponding generators) along with a residual gauge equivalence left over
from (\ref{18}):
\be
( \tilde{q}\; ,\; \tilde{p})\;\; \sim \;\; (W \tilde{q} W^{-1}\; ,\;
W \tilde{p} W^{-1})
= ( \tilde{q}'\; ,\; \tilde{p}') \; . \label{24}
\ee
$W$ here belongs to the Weyl subgroup of $G$. [For the case where $G$ is
$U(N)$ or $SU(N)$, this
action corresponds to a permutation of the diagonal entries of the
matrices corresponding to $ \tilde{q}$ and $ \tilde{p}$ in a representation
where the basis is chosen so that the $C^{i}$, the generators of
$\underline{C_{G}}$, are diagonal.]

(\ref{22}) and (\ref{23})
represent the free Hamiltonian for $n$ non-interacting particles on an
$n$-torus with some identifications ($n$ is the dimension of
$\underline{C_{G}}$ which equals $N-1$ for $\underline{SU(N)}$ and $N$ for
$\underline{U(N)}$). That this is so,  can be seen by defining variables
$\alpha
_{i}$:  \be
e^{-\alpha _{i}C^{i}} =  \tilde{q} \; . \label{25}
\ee
If $C^{i}$ are chosen such that
\be
e^{-L_{i} C^{i}}=1 \; \; \mbox{and} \; \;  e^{-\alpha_{i} C^{i}}\neq 1 \; \;
\mbox{for} \; \; 0<\alpha_{i} < L \; , \label{26}
\ee
then the $\alpha _{i}$ are coordinates on an $n$-torus taking values in the
range:
\be
0\leq \alpha _{i} \leq L_{i} \; . \label{27}
\ee

Now (\ref{24}) imposes further identifications on points of this torus.

In terms of $\alpha _{i}$ and $ \tilde{p}_{i}$, (\ref{22}) and (\ref{23}) are
\begin{eqnarray}
H & = & \pi \sum _{i}  \tilde{p}_{i}^{2} \; , \label{28} \\
\{ \alpha _{i}, \alpha _{j}\} = & 0 & = \{ \tilde{p}_{i},  \tilde{p}_{j}\} \;
, \nonumber \\
 \{ \alpha _{i},  \tilde{p}_{j}\} & = & \delta _{ij} \; .
\label{29} \end{eqnarray}
whereas (\ref{24}) can be rewritten as
\be
(\alpha _{i}\; ,\;  \tilde{p}_{i})\;\; \sim \;\; ((W\circ \alpha )_{i}\; ,\;
(W\circ \tilde{p})_{i}) \; , \label{30}
\ee
where $\circ$ is the action induced on the variables $\alpha_{i}$ and
$\tilde{p}_{i}$ by conjugation in the right hand side of (\ref{24}). (\ref{30})
is a discrete identification of the elements of the phase space arising because
of permutations of diagonal entries in the representation of $C^{i}$ mentioned
above.

The second method of quantizing pure YM on a cylinder now follows by taking
wavefunctions as complex square integrable functions of $\{ \alpha _{i}\} $,
thought of as coordinates of the $n$-torus which are however invariant under
(\ref{30}). More generally, the wavefunctions can be functions defined on the
universal cover of the $n$-torus which transform according to an irreducuble
unitary representation of the fundamental group of the configuration space,
compatibly with the condition (\ref{30}) lifted to the universal cover. In this
more general case, the wave functions can change by a
phase when transported around a non-trivial loop on the torus.

In the case of $SU(N)$ or $SU(N)/{\cal C}(SU(N))$, this more general case
trivializes because the equivalence (\ref{30}) forces the wave function to
remain unchanged when transported around one of the $N-1$ loops.

For $U(N)$, this happens for all the loops except the one corresponding
to its $U(1)$ center. Thus, in this case, wavefunctions can change by a phase
when they go around this particular loop while they are single-valued with
respect to the other $N-1$ loops.

The Hamiltonian (\ref{28}) is just the Laplacian in this representation and
so finding the eigenstates reduces to the easy problem of ``particle in a
box'', where along some directions [exactly one for $U(N)$ and none for $SU(N)$
or $SU(N)/{\cal C}(SU(N))$] wave functions are allowed to pick up a phase
around a closed loop.  Similar results, but using a somewhat different approach
have also been obtained in reference \cite{8}.

The third method of quantization is to go to the completely reduced phase space
which is ($T^{n}\times R^{n}$)/$\sim$ where $\sim$ is the equivalence under
(\ref{30}).  This quotienting however has fixed points and so the reduced phase
space becomes an orbifold.  Wave functions as before will be complex functions
of the reduced configuration space in the non-singular regions of the phase
space while the Hamiltonian will, as before, be the Laplacian in these regions.
The Laplacian will in general admit several self-adjoint extensions to the
singular regions (of necessarily lower dimension) and therefore will give
rise to a family of inequivalent quantum theories.

As an illustration of the
above general statements three specific examples will be considered below.

{\bf \underline{$G = SU(2)$}.}

This is a prototype of the $SU(N)$ case in which there are no $\theta$-states
and the reduced configuration space is also simply connected.

First, we have Rajeev's quantization \cite{2} according to which wavefunctions
are characters of irreducible representations. Since the Hamiltonian is
$\pi \,Tr(p^{2})$, the eigenvalues here are $2\pi j(j+1)$ where
$j=0,\frac{1}{2},1,\ldots$ .

In the second way of quantizing the system, we first observe that
$\underline{C_{G}}$ is 1-dimensional, consisting of the  the only element
$ \frac{1}{\sqrt{2}}\left( \begin{array}{cc}
                                           i & 0 \\
                                           0 & -i
                                           \end{array}
                                    \right)$ .

Eq. (\ref{28}) is now
\be
H=\pi \tilde{p}_{1}^{2} \; , \label{31}
\ee
while (\ref{29}) (in its quantum version) is
\begin{eqnarray}
{[ \alpha _{1},\alpha _{1}]}& = & 0\;\;\;=\;\;\;{[\tilde{p}_{1},\tilde{p}_{1}
]}
\nonumber \\
{[ \alpha _{1}, \tilde{p}_{1} ]}& = & i \; .\label{32}
\end{eqnarray}
Here $\alpha _{1}$ is the angular coordinate on a circle of length
$L=2\pi\sqrt{2}$.

(\ref{30}) leads to
\be
(\alpha _{1}\; ,\; \tilde{p}_{1})\;\;\sim \;\; (L -\alpha _{1}\; ,\;
-\tilde{p}_{1}) . \label{33}
\ee
as can be seen in the following way. Since
$\tilde{q}=\left( \begin{array}{cc} e^{-i\frac{\alpha _{1}}{\sqrt{2}}} & 0 \\
                                    0 & e^{i\frac{\alpha _{1}}{\sqrt{2}}}
\end{array} \right) $
while $\tilde{p}=\frac{1}{\sqrt{2}}\left( \begin{array}{cc}
i\tilde{p}_{1} & 0 \\
0 & -i\tilde{p}_{1}
\end{array} \right)$, the permutation that $W$ causes, takes $\tilde{q}$
to $\tilde{q}'=\left( \begin{array}{cc}
e^{i\frac{\alpha _{1}}{\sqrt{2}}} & 0 \\
0 & e^{-i\frac{\alpha _{1}}{\sqrt{2}}}
\end{array} \right) $ and $\tilde{p}$ to $\tilde{p}'=\frac{1}{\sqrt{2}}\left(
\begin{array}{cc} -i\tilde{p}_{1} & 0 \\
0 & i\tilde{p}_{1}
\end{array} \right) $.  Hence (\ref{33}) follows.

Thus the second method of quantization would correspond to considering
wave functions $\Psi(\alpha _{1})$ in the Hilbert space of square
integrable functions on $S^{1}$ which are invariant under:
\be
\alpha _{1} \;\; \longrightarrow \;\; L -\alpha _{1} \; . \label{34}
\ee
[The more general wavefunction which may transform by a phase under $\alpha
_{1} \;\;\longrightarrow \;\; L +\alpha _{1}$ is automatically excluded
because of the requirement that $\Psi (0)=\Psi (L)$  arising from
(\ref{34}).]

A complete set of functions on a circle which are also eigenfunctions of the
Hamiltonian (\ref{31}) (which in this representation is just given by the
operator $-\pi \frac{d^{2}}{d\alpha _{1}^{2}}$) are
$\Psi(\alpha_{1})=\sqrt{\frac{1}{L}} e^{i\frac{2\pi}{L}
n\alpha _{1}}\;\;n\in \bf{Z}$.
However, with restriction (\ref{34}), the set of
functions allowed are only
\begin{center}
$\cos \frac{2\pi}{L} n \alpha_{1} \;\; , \; \; n \in {\bf Z}
\;.$ \end{center}
Thus the eigenfunctions are
\be
\Psi _{n}(\alpha _{1}) = \sqrt{\frac{2}{L}} \cos \frac{2\pi}{L}n\alpha
_{1} \; , \label{35} \ee
while the corresponding eigenvalues are
\be
E_{n} = \frac{\pi n^{2}}{2} \; . \label{36}
\ee

The third way of quantizing $SU(2)$ YM on a cylinder is by doing the quotient
by the equivalence relation (\ref{33}). It yields the segment
${[0,\frac{L}{2}]}$ as the configuration space. The corresponding phase space
is
given by ${(\tilde{q}\; ,\; \tilde{p}) \in [0\; ,\; \frac{L}{2} ]\times
\bf{R}}$
with the identifications $(0,\tilde{p}_{1}) \equiv (0,-\tilde{p}_{1})$ and
$(\frac{L}{2}, \tilde{p}_{1}) \equiv (\frac{L}{2} , -\tilde{p}_{1})$.  Thus the
interior of the phase space is $(0,\frac{L}{2})\times \bf{R}$ and the
interior of the  coordinate space is the open interval $(0,\frac{L}{2})$.
Here the
wavefunctions can therefore be considered as functions of $x\in
[0,\frac{L}{2}]$ while
the Hamiltonian $H$ is any self-adjoint extension of $ -\pi
\frac{d^{2}}{d\alpha_{1}^{2}}$ from the interior $(0,\frac{L}{2})$ to the
closed segment,
obtained by imposing suitable boundary conditions.

Let us recall that an operator $A$ is self-adjoint on a domain ${\cal D}$ if
\begin{center}
$<\Psi |A \Phi> = <A \Psi | \Phi > \;\;\;\; \forall \Phi \in {\cal D}
\Leftrightarrow \Psi \in {\cal D}$ ,
\end{center}
$<\cdot|\cdot>$ indicating the scalar product. For
$A=H=-\frac{d^{2}}{d\alpha_{1}^{2}}$ on $[0,\frac{L}{2}]$, one can easily show
that
the above condition is equivalent to find a domain ${\cal D}\subset
L^{2}([0,\frac{L}{2}])$ which satisfies:
\begin{center}
$\frac{d}{d\alpha_{1}}\bar{\Psi}(\alpha_{1}) \, \Phi(\alpha_{1}) -
\bar{\Psi}(\alpha_{1}) \, \frac{d}{d\alpha_{1}}\Phi(\alpha_{1})\,
|^{L/2}_{0} = 0 \; \;  \forall \Phi \in {\cal D} \Leftrightarrow \Phi \in
{\cal D}$ . \end{center}
There is an infinite nunber of such domains \cite{7}, corresponding to
different boundary conditions for the wavefunction $\Psi (\alpha_{1})$ at
$\alpha_{1}=0,\frac{L}{2}$.

The spectrum and eigenfunctions of $H$ depend on the extension chosen.

For example, for the self-adjoint extension corresponding to the boundary
condition $\frac{d}{d\alpha_{1}}\Psi(0)=
\frac{d}{d\alpha_{1}}\Psi(\frac{L}{2})=0$, one
recovers the set of eigenfunctions  and
eigenvalues (\ref{35}) and (\ref{36}).

If the boundary condition reads $\Psi(0)=\Psi(\frac{L}{2})=0$ instead, the
eigenfunctions are
\be
\Psi _{n}(\alpha _{1}) = \sqrt{\frac{4}{L}} \sin \frac{2\pi}{L} n\alpha _{1} \;
,
\label{38} \ee
whereas the eigenvalues are still given by
\be
E_{n} = \frac{\pi n^{2}}{2} \; . \label{39}
\ee

On the contrary, the eigenvalue do change for other boundary conditions, such
as
$\Psi(\frac{L}{2})=e^{i\theta} \, \Psi(0)$ ,
$\frac{d}{d\alpha_{1}}\Psi(\frac{L}{2})=e^{i\theta}
 \, \frac{d}{d\alpha_{1}}\Psi(0)$ . In this case we have:
\begin{eqnarray}
\Psi _{n}(\alpha _{1}) & = & \sqrt{\frac{2}{L}}
e^{i\frac{2\pi}{L}(2n+\frac{\theta}{\pi})\alpha _{1}} \; , \label{40}   \\
E_{n} & = & \frac{\pi (2n+\frac{\theta}{\pi})^{2}}{2} \; . \label{41}
\end{eqnarray}
(\ref{36}) and (\ref{41}) are not the same, nor do they agree with the spectrum
found by Rajeev.

{\bf \underline{$G$ = $SO(3)$}.}

This is the prototype of the $SU(N)/{\cal C}(SU(N))$ case, where there are
$\theta$-states but the reduced configuration space is simply connected so that
the usual relation between them is violated. This example is very similar to
the
previous one except that the group is now doubly connected.  This means that
the
analogue of $q$ here when written in terms of an $SU(2)$ group element is the
equivalence class \{ $\pm q$\} ($-{\bf 1}$ is the element of $SU(2)$ which
also maps to the identity of $SO(3)$).

Firstly, note that we have the relation
\begin{flushright}
$(q\; ,\; p)\;\; \sim \;\; (gqg^{-1}\; ,\; gpg^{-1})$ \hspace{4cm} (18)
\end{flushright}
where $q,g \in SU(2)$ and $p \in \underline{SU(2)}$.

(\ref{18}) gives the gauge transformations that are generated by Gauss law and
therefore wave functions have to be invariant under these transformations.

There are also gauge transformations not connected to
identity, which are not generated by Gauss law. Wave functions
can transform under unitary representations of this group of
transformations, this less stringent requirement being enough to guarantee
invariance of the matrix elements of observables. In this example $g(x)$
such that $g(0) = {\bf 1}$ and $g(2\pi )=-{\bf 1}$ is a gauge
transformations that is not connected to identity. The equivalence that it
causes is:
\be
(q,p) \; \; \sim \; \; (-q,p) \; . \label{42}
\ee
In terms of the variables defined in (\ref{21}), this means that
\be
(\tilde{q}\; ,\; \tilde{p})\; \; \sim \; \; (-\tilde{q}\; ,\; \tilde{p}) \; ,
\label{43}
\ee
or in terms of the $\alpha _{1}$ and $p_{1}$ defined in (\ref{25}) and
(\ref{3})
respectively,
\be
(\alpha _{1}\; ,\; \tilde{p}_{1})\; \; \sim \; \; (\alpha _{1}+\frac{L}{2}\;
,\;
\tilde{p}_{1}) \;. \label{44}
\ee
The corresponding group of transformations here is ${\bf Z}_{2}$.
Thus wave functions transform under unitary representations of ${\bf Z}_{2}$:
\be
\Psi (\alpha _{1}+\frac{L}{2} )=\pm \Psi (\alpha _{1}) \; . \label{45}
\ee
This is just the usual argument given for the existence of $\theta$-states.

In Rajeev's method of quantization, since wave functions are taken as
characters of irreducible representations of $SU(2)$, the representations
corresponding to $\frac{1}{2}$-integers satisfy $\Psi (-q)=-\Psi (q)$ while
the integral representations satisfy $\Psi (-q)=\Psi (q)$. They correspond to
the two choices in (\ref{45}). Clearly we cannot allow wavefunctions of both
types to occur in a single quantum theory because then a superposition of these
wave functions would not satisfy (\ref{45}). Thus two quantum theories
are super-selected, one for each sign in (\ref{45}). The energies for the
quantum theories corresponding to the $+$ sign and $-$ sign are $E_{j}=2\pi
j(j+1)$ with $j$ being integer or half-integer respectively.

In the second method of quantization, eigenfunctions are as in (\ref{35})
above.
With condition (\ref{45}) there is again  a super-selection into two quantum
theories with eigenfunctions
\begin{eqnarray}
\Psi _{n}^{I}(\alpha _{1})=\sqrt{\frac{4}{L}} \cos \frac{2\pi}{L} (2n) \alpha
_{1} & \mbox{for} & \Psi (\alpha _{1}+\frac{L}{2} )=\Psi (\alpha _{1}) \; ,
\nonumber \\  \Psi_{n}^{I\!\!I}(\alpha _{1})= \sqrt{\frac{4}{L}} \cos
\frac{2\pi}{L}(2n+1)\alpha _{1} & \mbox{for} &  \Psi(\alpha _{1}+\frac{L}{2})
=-\Psi (\alpha _{1})
\label{46}  \end{eqnarray}
and the corresponding energy eigenvalues
\begin{eqnarray}
E_{n}^{I} & = & \frac{\pi (2m)^{2}}{2} \; , \nonumber \\
E_{n}^{I\!\!I} & = & \frac{\pi (2m+1)^{2}}{2} \; . \label{47}
\end{eqnarray}

In the third way of quantization, we have to quotient first by (\ref{30}) to
get the
completely reduced phase space as in the $SU(2)$ example. In addition we have
relation (\ref{44}) and the corresponding rule for wave functions (\ref{45}) to
contend with. In the reduced phase space, (\ref{44}) and (\ref{45})
will read
\begin{eqnarray}
(\alpha _{1}\; ,\; \tilde{p}_{1}) & \sim & (\frac{L}{2}
-\alpha _{1}\; ,\; -\tilde{p}_{1}) \; , \label{48} \\
\Psi (\frac{L}{2} -\alpha _{1}) & = &
\pm \Psi (\alpha _{1}) \; . \label{49} \end{eqnarray}

In this case,while the boundary conditions at $\alpha_{1}=0$ are arbitrary,
those at $\alpha_{1}=\frac{L}{4}$ are not. This is so because the reduced
configuration space is (as before) the interval $[0,\frac{L}{2}]$, but now
condition
(\ref{49}) forces $\frac{d}{d\alpha_{1}}\Psi(\frac{L}{4})$ or
$\Psi(\frac{L}{4})$ to be zero depending on the $+$ or $-$ choices of
(\ref{49}). Thus the problem reduces to finding those self-adjoint extensions
of $\frac{d^{2}}{d\alpha_{1}^{2}}$ on the interval $[0,\frac{L}{2}]$ such that
either
$\frac{d}{d\alpha_{1}}\Psi(\frac{L}{4})=0$ or $\Psi(\frac{L}{4})=0$. Here,
the most general boundary condition at $\alpha_{1}=0$ for
$\frac{d^{2}}{d\alpha_{1}^{2}}$ to be self-adjoint is
\be
\frac{d}{d\alpha_{1}}\Psi(0) = k \Psi(0) \; , \label{50}
\ee
where $k$ is an arbitrary real constant.

The energy levels are dependent on the parameter $k$ and whether the BC at
$\alpha_{1}=\frac{L}{4}$ is the one corresponding to the $+$ or $-$ sign of
(\ref{49}). For example, when $k=0$, the eigenstates and eigenvalues reduce to
those obtained in (\ref{46}) and (\ref{47}) above.

Incidentally, (\ref{48}) also shows that the completely reduced configuration
space is just a segment (the closed interval $[0,\frac{L}{4}]$) and hence is
simply connected. So, the argument which relates $\theta$-states to the
multiple
connectivity of the configuration space clearly breaks down here since
$\theta$-states do exist despite the configuration space being simply
connected.

One further point to be noted here is the following. If instead of sticking to
condition (\ref{49}), we treat (\ref{48}) as another exact equivalence and
quotient the phase space by this equivalence, we arrive at a completely reduced
phase space identical in structure to that of the $SU(2)$ example, save for the
fact that the coordinate variable $\alpha_{1}$ here takes values in
$[0,\frac{L}{4}]$ instead of $[0,\frac{L}{2}]$. So, the eigenfunctions and
eigenvalues obtained in this approach match those obtained in the third way of
quantizing $SU(2)$ YM [Equations (37-41)] with the following replacements:
\begin{eqnarray}
\alpha_{1} & \rightarrow & 2\alpha_{1} \; , \nonumber \\
E_{n} & \rightarrow & 4E_{n} \; . \label{51}
\end{eqnarray}
In this approach the BC at $\alpha_{1}=\frac{L}{4}$ has been relaxed and a
two-fold ambiguity [$\frac{d}{d\alpha_{1}}\Psi(\frac{L}{4})=0$ or
$\Psi(\frac{L}{4})=0$] has been allowed to become an $\infty$-fold
ambiguity. This infinity of BC's now also contains those corresponding to
$\theta$-states as special cases.

{\bf \underline{$G$ = $U(2)$}.}

This is a prototype of the $U(N)$ case in which $\theta$-states exist and are
also related to the fundamental group of the reduced configuration space in the
standard way. In this example, there are gauge transformations not connected to
the identity with the added feature that now they act freely on ${\cal A}/{\cal
G}_{0}$, so that ${\cal G}/{\cal G}_{0}$ acts freely on ${\cal A}/{\cal
G}_{0}$. Thus the reduced configuration space is
\be
{\cal Q} = {\cal A}/{\cal G} = ({\cal A}/{\cal G}_{0})/({\cal G}/{\cal G}_{0})
\label{52}
\ee
and therefore
\be
\pi_{1}({\cal Q}) \equiv \pi_{1}({\cal A}/{\cal G}) = \pi_{0}({\cal G}/{\cal
G}_{0}) = {\cal G}/{\cal G}_{0} = {\bf Z} \; . \label{53}
\ee
Therefore $\theta$-states here are related to the first homotopy group
$\pi_{1}({\cal
Q})$ of the reduced configuration space and the former can be thought of as
arising because of quantization ambiguities corresponding to different unitary
representations of $\pi_{1}({\cal Q})={\bf Z}$.

If $U(2)$ is parametrized using an $U(1)$ element $u$ and an $SU(2)$ element
$q$, then an element of $U(2)$ corresponds to the pair $(u,q)$ provided the
following identification is made:
\be
(u,q) \sim (-u,-q) \; . \label {54}
\ee

Thus wave functions in Rajeev's approach are complex functions of the above
pair which are such that
\begin{eqnarray}
\Psi[(u,gqg^{-1})] & = & \Psi[(u,q)] \; , \nonumber \\
\Psi[(-u,-q)] & = & \Psi[(u,q)] \, e^{i\theta} \; . \label{55}
\end{eqnarray}
The first condition simply restates the similar condition imposed on wave
functions even in the $SU(2)$ case. The second condition is a consequence of
(\ref{54}) and the multiple connectivity of $U(2)$ which allows wave functions
to transform under a unitary representation of the fundamental group (here
${\bf Z}$) when the coordinate variable goes around a non-trivial
loop.

The Hamiltonian in Rajeev's method is the quadratic Casimir of $U(2)$ and the
eigenstates are the $SU(2)$ characters multiplied by functions of one
coordinate (parametrizing the $U(1)$ center) which are of the form (\ref{40}).
More specifically, if $\psi_{n}(\alpha)=e^{i(n+\frac{\theta}{2\pi})\alpha}$
and $\chi$ is one of the characters of $SU(2)$, then the eigenstates are
\be
\Psi[(u,q)] = \psi_{n}(\alpha) \chi(q) \; , \label{56}
\ee
where $u=e^{i\alpha}$ and $n=2m+1$ or $n=2m$ depending on $\chi$ being a
character of a half-integer or integer representation respectively. The energy
eigenvalues are \be
E_{j,n} = 2 \pi j (j+1) + \pi (n+\frac{\theta}{2\pi})^{2} \; , \label{57}
\ee
where $j$ takes values $0,\frac{1}{2},1,\ldots$ and $n$ takes values
$0,1,2,\ldots$ .

The inequivalent quantizations here correspond to different values of the
parameter $\theta$ and they can be thought of as arising due to the existence
of gauge transformations not connected to the identity ($\theta$-states
approach) or equivalently be thought of as arising due the existence of a
non-trivial fundamental group for the reduced configuration space.

Similar analysis (not repeated here) goes through for the other approaches to
quantization. In each of the other approaches too, an extra term of the form
(\ref{40}) multiplies the already existing $SU(2)$ eigenstates while a
corresponding extra term of the form (\ref{41}) adds to the energy eigenvalues
of the $SU(2)$ example.
\vskip2cm

\begin{large}
{\bf Acknowledgements}
\end{large}

We thank A.P. Balachandran for encouraging us to look into this problem and
providing many useful suggestions. We thank also Arshad Momen, Giuseppe Bimonte
and Paulo Teotonio-Sobrinho for several fruitful discussions.

This work was supported in part by the Department of Energy, U.S.A., under
contract number DE-FG02-85ER40231.

\end{document}